\documentclass[journal]{IEEEtran}
\usepackage{blindtext}
\usepackage{graphicx}
\usepackage{textcomp}
\usepackage{fancyhdr}
\usepackage{cite}
\usepackage{booktabs}
\usepackage{caption}

\begin{document}

%
\title{Color-based classification of EEG Signals for people with severe locomotive disorder}
%
%
%
\author{
  Shrestha, Ankit
  (\texttt{ankit.shrestha@usu.edu})
  \and
  Adhikari, Bikram
  (\texttt{badhika5@gmu.edu})
}

%
%

\markboth{Color based classification of EEG Signals for people with severe locomotive disorder}%
{Shell \MakeLowercase{\textit{et al.}}: Bare Demo of IEEEtran.cls for Journals}


\maketitle
\fancypagestyle{plain}{
\fancyhf{}         
\fancyfoot[L]{}
\fancyfoot[C]{}
\fancyfoot[R]{}
\renewcommand{\headrulewidth}{0pt}
\renewcommand{\footrulewidth}{0pt}
}
\pagestyle{fancy}{
\fancyhf{}
\fancyfoot[R]{}}
\renewcommand{\headrulewidth}{0pt}
\renewcommand{\footrulewidth}{0pt}

\begin{abstract}
The neurons in the brain produces electric signals and a collective firing of these electric signals gives rise to brainwaves. These brainwave signals are captured using EEG (Electroencephalogram) devices as micro voltages. These sequence of signals captured by EEG sensors have embedded features in them that can be used for classification. The signals can be used as an alternative input for people suffering from severe locomotive disorder.Classification of different colors can be mapped for many functions like directional movement. In this paper, raw EEG signals from NeuroSky Mindwave headset (a single electrode EEG sensor) have been classified with an attention based Deep Learning Network. Attention based LSTM Networks have been implemented for classification of two different colors and four different colors. An accuracy of 93.5\% was obtained for classification of two colors and an accuracy of 65.75\% was obtained for classifcation of four signals using the mentioned attention based LSTM network.
\end{abstract}


\IEEEpeerreviewmaketitle{}

\section{Introduction}
According to WHO, 15\% (around 1 billion) people of total world population lives with various form of disability. Among which 2-4\% are suffering from severe form of disability \cite{who2011world}.\par
Many researches are carried out to improve the quality of life of people with disabilities. Among many, Brain Computer Interface (BCI) is one of the highly researched. BCI is a technology where a person can control the external device directly through thought process. Electroencephalography (EEG) is one of the widely used method to track brain wave patterns for BCI. Various devices are developed to help disabled people in day to day activity like smart wheelchair \cite{5354534}, communication system for people with lock-in syndrome \cite{6151455}, and 
using EEG and BCI for communication.\cite{wolpaw_2004}\par
As of 2018, with a consumer toll of 3.3 million wheelchair users alone in USA, of which 1.825 of the age 65 and above,\cite{kdsmartchair} there has been a lot of research with making a wheelchair more accessible and easy to use and a lot of the research was focused on finding out ways for people with wheelchair to use it without any assistance.\par
With EEG being one of the areas researched upon, this research will be mainly focused on the application of BCI and EEG signal classification to improve the quality of life of people with disabilities. This paper explains how BCI and EEG in sync help affected person to control locomotion and communication on their own.
\section{Literature review}\par
BCI-EEG has become a huge field of research in the present days in order to augment the brain through computer. A wide range of BCI-EEG application have been researched and developed such as mind controlled robots, smart environment control, etc.\par
A BCI for controlling a lower limb exoskeleton was developed by the scientists from Korea University and TU Berlin\cite{article1} with the help of EEG cap to control forward, left and right movement as well as to stand and sit by simply staring at flickering light emitting diodes (LEDs). This application used steady-state visual potential (SSVEP) approach along with canonical correlation analysis (CCA) method in combination with k-nearest neighbors.\par
Another approach to BCI was to control Smart Home Environment using Conceptual Imagery 
\cite{article2}, where the semantics of the interaction is compatible with the semantic of the task. For e.g., when the user imagine the concept of a TV set then BCI recognizes the concept and generates  a command to turn on the TV set for the user to watch. This concept has been experimented by Nataliya Kosmyna and her team with the help of “Domus”, a fully functional 40 meter square smart home with a bedroom, a kitchen, a bathroom and a living room. They used an asynchronous BCI system based on a Minimum Distance Classifier (MDCs) that produced an accuracy in task of 77\% in healthy subjects and a better accuracy of 81\% in disabled subjects.
\par
Puja Sengupta proposed a Continuous Monitoring System~\cite{moultonr_2019}, for analyzing the Locked-In Syndrome(LIS) Patients and Comatose Patients to help them be diagnosed properly by continuously monitoring the patient. The system used Emotiv EpocPlus Headset to collect the EEG signals of the Coma patient and the signals were classified using a Multilayer Feedforward Neural Network including Restricted Boltzmann Machine(RBM) algorithm in the first hidden layer and Sigmoid function and Radial Basis Function(RBF) algorithm in the second hidden layer.
\par
Intelligent adaptive user interface (iAUI) was used in a EEG Based Mobile Remote Control for Wheel Chair Navigation through adaptive brain-robot interface~\cite{gandhi2014eeg}.\par
Another approach may be through intent recognition~\cite{zhang2017intent}, which can be used for smart applications such as turning off the led or controlling a robot. 
LSTM~\cite{hochreiter1997long} tuned using implementation of Orthogonal Array experiment, they were able to achieve considerably higher accuracy.\par
This paper describes the implementation of BCI in the medical field to help people with disabilities by researching on the development of an application that is capable of mind-controlled wheelchair movement and detection of coma patients’ thoughts. Neurosky Mindwave has been used for the collection of EEG data and Deep learning approach has been used for the classification of the data obtained. With the implementation of Recurrent Neural Network (RNN) and Long Short Term Memory (LSTM) neural network along with attention layer, there has been a significant improvement in the results obtained in comparison to the previous approaches used.\par
\section{Methodology}

\subsection{Data Acquisition and Pre-processing}

Neurosky Mindwave, a single electrode EEG headset, has been used for the purpose of data collection for this project. The collection of data was carried out on unstressed and natural conditions with minimal interference and noise from the environment.\par
The data was collected at 512 Hz which is the maximum frequency that this EEG headset supports. Four different colors were used as stimuli for the input signal to EEG: red, blue, black and yellow. A corresponding label/command of 0, 1, 2 and 3 were mapped with the colors, respectively.\par
The subjects were instructed to first input the particular color that they are going to think of and then focus on their thoughts for the next four seconds. The headset captured the data for 4 seconds resulting in a sequence of 2048 numbers. The data with its attention level and signal quality were then stored along with the input label.\par
The details are given in table~\ref{tab:command}.
\begin{table}
    \centering
    \caption{Command for signals}
\label{tab:command}
    \begin{tabular}{p{0.2in} p{0.2in} p{0.2in}}
        \toprule
        Action&Color&Command\\
        \midrule
        left&red&0\\
        right&blue&1\\
        up&black&2\\
        down&yellow&3 \\\bottomrule
    \end{tabular}
\end{table}\par
During data collection, the obtained signals were filtered for outliers and abnormal/error signals to minimize noise in time of input. The data was then introduced to shape changes for feeding to neural network. The shape was changed from a long sequence of 2048 to a shape of (256, 8). Further preprocessing was not done and the raw data was used for classification in the mentioned state.
\subsection{Classification Model}
Deep learning approaches were applied for the classification of the EEG signals. Recurrent Neural Networks (RNNs) were used in the neural network architecture as the RNNs can preserve the dynamic temporal behavior of the sequences. Many variations and types of RNNs were tried to realize the deep neural net for classification. After many trials, LSTM network with an attention layer was chosen as the best fit for this problem.\par
An implementation of LSTM neural network with attention layer was carried out to classify the signals. Two LSTM layers with 256 memory units and 64 memory units respectively were stacked for this neural network-based classifier. The first LSTM layer was designed to input sequences of shape (512,4) and was then connected to the dropout layer which was activated using Leaky-RELU. This was connected to the second LSTM layer with 64 memory units which was connected to an attention layer that learnt to focus on the important features of the signal. The output of the attention layer was then fed to the dense layer with SoftMax activation which acted as the output layer for the classifier. Adam algorithm, an extension of the stochastic gradient descent (SGD), was used as the optimizer for the classifier. Cross Entropy was used as the loss function for the classifier. Since, this was a multiclass problem, Categorical Cross Entropy was used as the loss function.\par
The same model of the neural network was trained using different class of data. A four-class classifier was implemented for applications like Direction control of Wheelchair and a two-class classifier was implemented for applications like Detection of Coma patient thought for Yes/No questions. For the two-class classifier the two colors with the best accuracy was used as input after multiple trials.\par
A brief introduction to the network models are given in the following sections.
\subsubsection{RNN}
A recurrent neural network (RNN) is a class of artificial neural network where connections between nodes form a directed graph along a sequence. This allows it to exhibit dynamic temporal behavior for a time sequence. Unlike feed-forward neural networks, RNNs can use their internal state (memory) to process sequences of inputs.~\cite{wiki:xxx}\par
\subsubsection{Bi-directional RNN}
Bi-directional RNNs use a finite sequence to predict or label each element of the sequence based on the element's past and future contexts. This is done by concatenating the outputs of two RNNs, one processing the sequence from left to right, the other one from right to left. The combined outputs are the predictions of the teacher-given target signals. This technique has proved to be especially useful when combined with LSTM RNNs.\par
\begin{figure}
    \centering
    \captionsetup{justification=centering}
    \includegraphics[width=3cm]{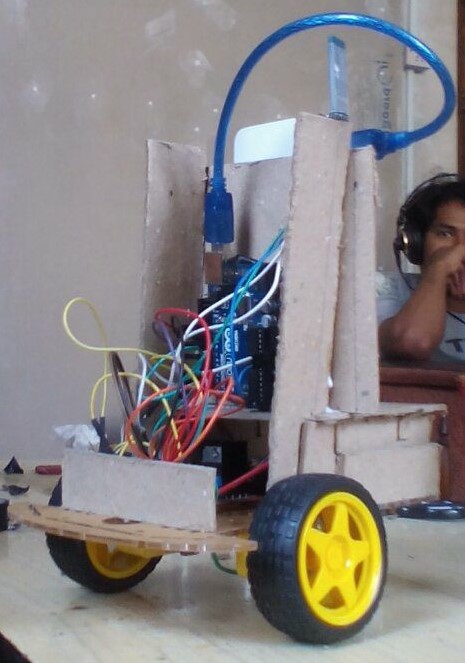}
    \caption{The actual prototype for the wheelchair implemented for the research}
\label{fig:model}
\end{figure}

\begin{figure}
    \centering
    \captionsetup{justification=centering}
    \includegraphics[width=4cm]{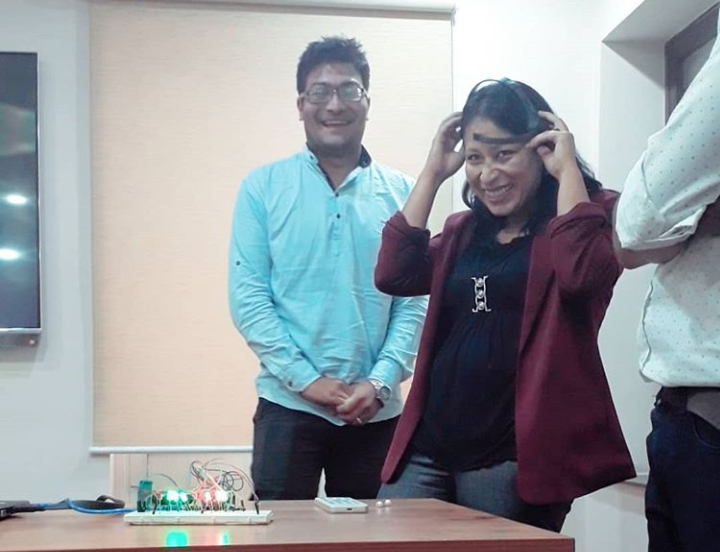}
    \caption{The yes/no question response simulator prototype being tested among test subjects}
\label{fig:subject}
\end{figure}

\subsubsection{Long short-term memory (LSTM)}
Long short-term memory (LSTM) ~\cite{Hochreiter:1997:LSM:1246443.1246450} is a deep learning system that avoids the vanishing gradient problem. LSTM is normally augmented by recurrent gates called ``forget'' gates. LSTM prevents back propagated errors from vanishing or exploding. Instead, errors can flow backwards through unlimited numbers of virtual layers unfolded in space. That is, LSTM can learn tasks that require memories of events that happened thousands or even millions of discrete time steps earlier.\par
\subsubsection{Attention Network}
An attention function can be described as mapping a query and a set of key-value pairs to an output, where the query, keys, values, and output are all vectors.~\cite{attentionisallyouneed} The output is computed as a weighted sum of the values, where the weight assigned to each value is computed by a compatibility function of the query with the corresponding key\par
Now for the wheelchair, a simple raspberry 3.0 powered wheelchair with two motors for two side wheel was used to manure the wheelchair. The table~\ref{tab:command}. demonstrates the motor state as per the signal received:
\begin{table}[h!tpb]
    \centering
    \caption{Command for Wheelchair and the motor state}
    \begin{tabular}{p{0.2in} p{0.8in} p{0.8in} p{0.4in}}
        \toprule
        Signal&Motor1&Motor2&Action\\ \midrule
        0&Anti-clockwise&Anti-clockwise&Back\\
        1&Clockwise&Clockwise&Front\\
        2&Off&Clockwise&Left\\
        3&Clockwise&Off&Right\\\bottomrule
    \end{tabular}
\end{table}\\
 and the figure: \ref{fig:model} refrences the prototype model being implemented for the four channel classifier.\par
\begin{figure}
    \centering
    \captionsetup{justification=centering}
    \includegraphics[width=6cm]{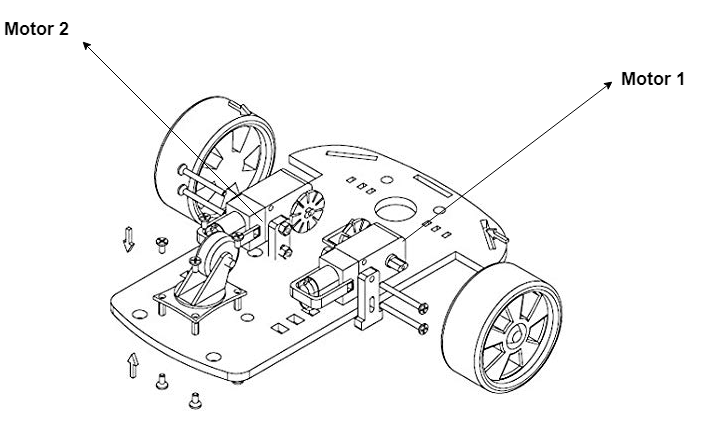}
    \caption{The wireframe design for the two motor based programmable car chasis used as a wheelchair prototype.}
\label{fig:a2}
\end{figure}

\section{Result and Analysis}
\subsection{Specification of System}
Keras framework with tensorflow was used to build this model. It was trained on 1050 TI 4 GB memory, I5 8th gen 8300H with 8 GB ram. For the wheelchair prototype, an arduino uno, raspberry pi 3.0 and a programmable car chasis set with two motors for left and right wheels were used. 

\subsection{Approach}

The system was trained using the EEG signals for four colors (red-blue-black-yellow). The model was trained using a complete data set, with a validation split to obtain the point after which the data starts over-fitting in the network. After this, the training data at hand were again fed to train the neural network model upto the point where the model is most appropriately fit. The EEG data set used in the test were recorded for four colors and contained both trained and untrained data.\par
 Of 200 EEG signals collected from numerous subjects that were used for training the system, 50 seen and 50 unseen data were used to test for each color. For a four class classifier, the accuracy obtained for each direction were combined to give out the total accuracy, whereas for two class classifier, each color with an adjacent color were trained and the combination showing the best accuracy based on trial was used for further processing.
\begin{figure}
    \centering
    \captionsetup{justification=centering}
    \includegraphics[width=6cm]{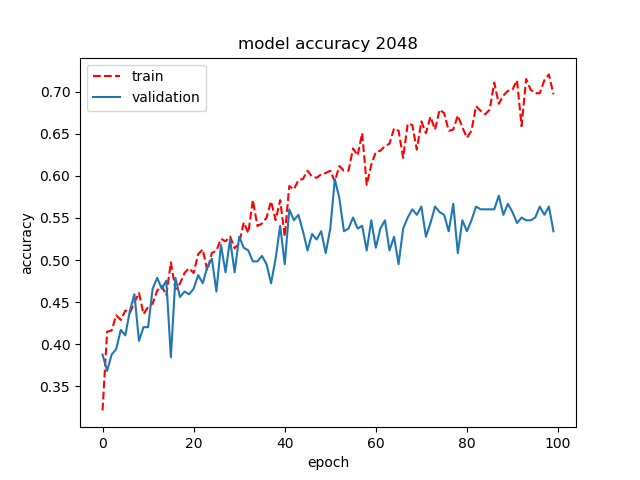}
    \caption{The graph shows the training and validation accuracy obtained between all four colors over 100 epochs.}
\label{fig:allacc}
\end{figure}
\begin{figure}
    \centering
    \captionsetup{justification=centering}
    \includegraphics[width=6cm]{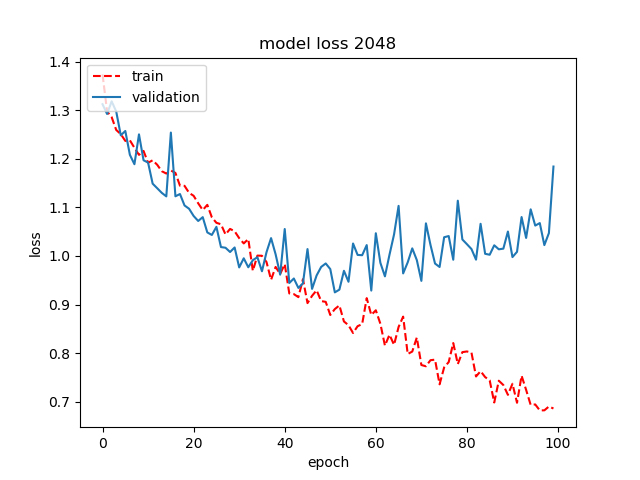}
    \caption{The graph shows the training and validation loss obtained between all four colors over 100 epochs.}
\label{fig:allloss}
\end{figure}
\begin{figure}
    \centering
    \captionsetup{justification=centering}
    \includegraphics[width=6cm]{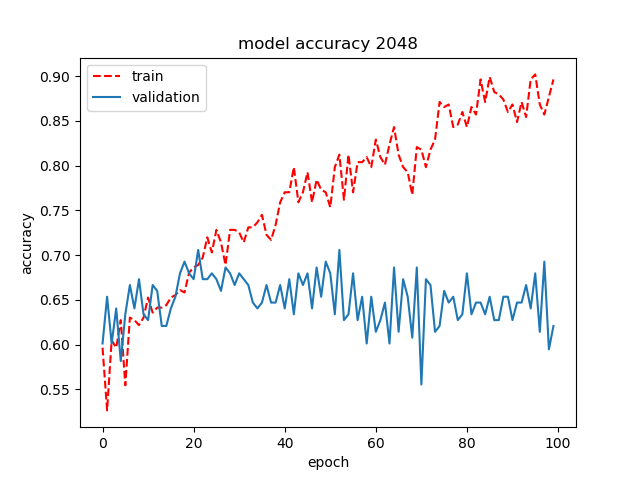}
    \caption{The graph shows the training and validation accuracy obtained between red and black over 100 epochs.}
\label{fig:acc01}
\end{figure}
\begin{figure}
    \centering
    \captionsetup{justification=centering}
    \includegraphics[width=6cm]{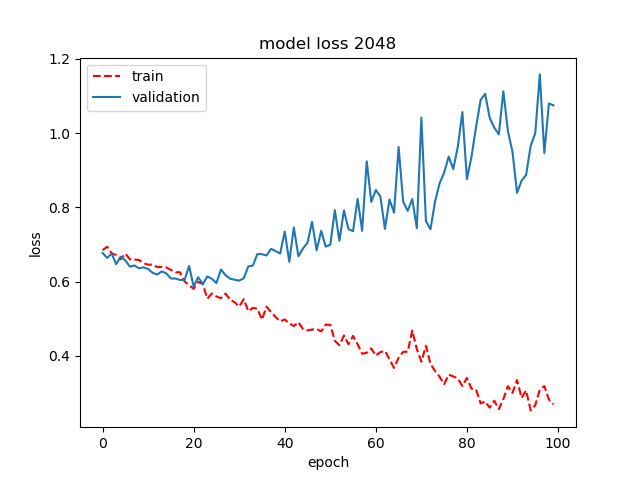}
    \caption{The graph shows the training and validation loss obtained between red and black over 100 epochs.}
\label{fig:loss01}
\end{figure}
\begin{figure}
    \centering
    \captionsetup{justification=centering}
    \includegraphics[width=6cm]{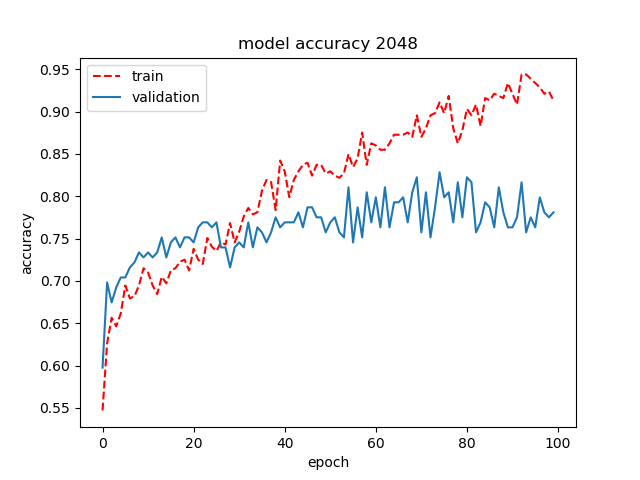}
    \caption{The graph shows the training and validation accuracy obtained between red and blue over 100 epochs.}
\label{fig:acc02}
\end{figure}
\begin{figure}
    \centering
    \captionsetup{justification=centering}
    \includegraphics[width=6cm]{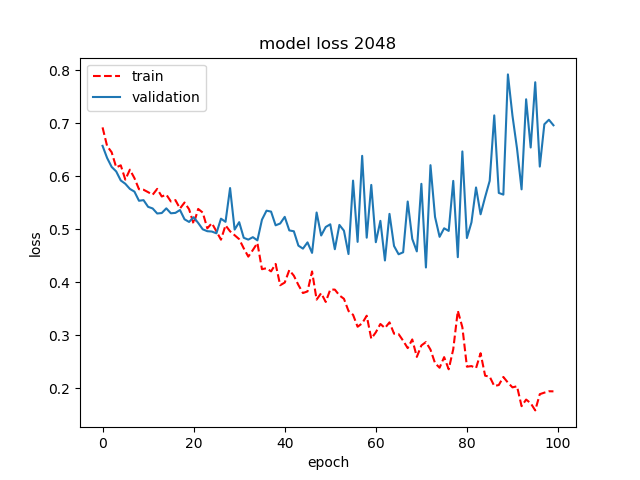}
    \caption{The graph shows the training and validation loss obtained between red and blue over 100 epochs.}
\label{fig:loss02}
\end{figure}
\begin{figure}
    \centering
    \captionsetup{justification=centering}
    \includegraphics[width=6cm]{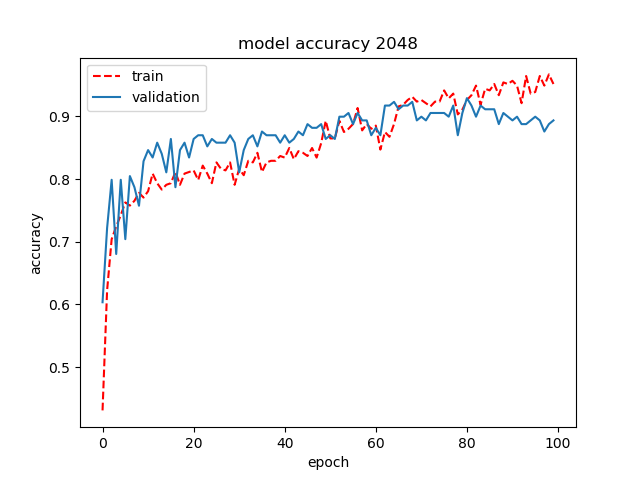}
    \caption{The graph shows the training and validation accuracy obtained between red and yellow over 100 epochs.}
\label{fig:acc03}
\end{figure}
\begin{figure}
    \centering
    \captionsetup{justification=centering}
    \includegraphics[width=6cm]{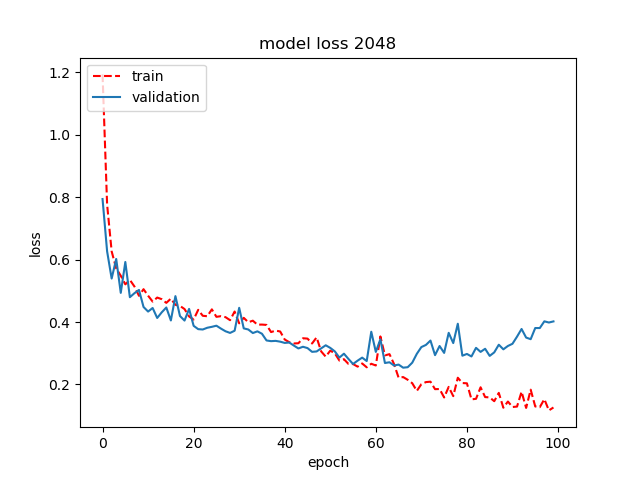}
    \caption{The graph shows the training and validation loss obtained between red and yellow over 100 epochs.}
\label{fig:loss03}
\end{figure}
\begin{figure}
    \centering
    \captionsetup{justification=centering}
    \includegraphics[width=6cm]{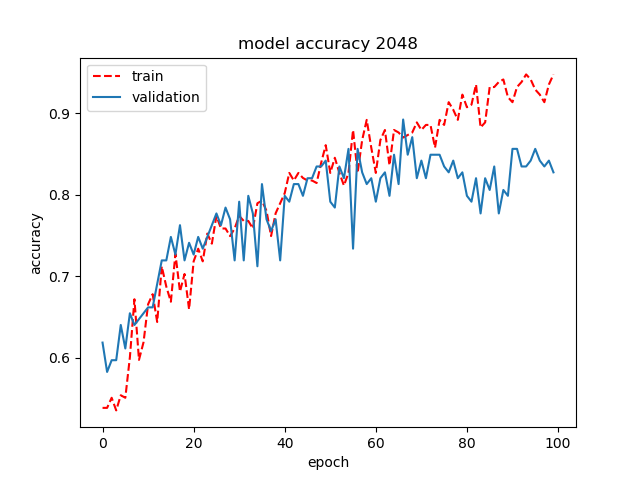}
    \caption{The graph shows the training and validation accuracy obtained between black and blue over 100 epochs.}
\label{fig:acc12}
\end{figure}
\begin{figure}
    \centering
    \captionsetup{justification=centering}
    \includegraphics[width=6cm]{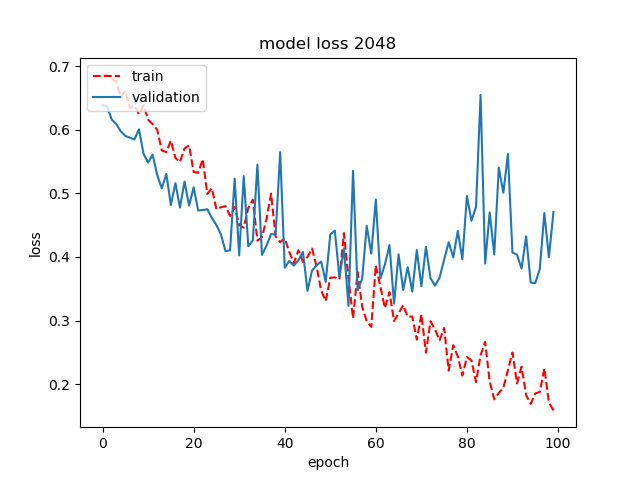}
    \caption{The graph shows the training and validation loss obtained between black and blue over 100 epochs.}
\label{fig:loss12}
\end{figure}
\begin{figure}
    \centering
    \captionsetup{justification=centering}
    \includegraphics[width=6cm]{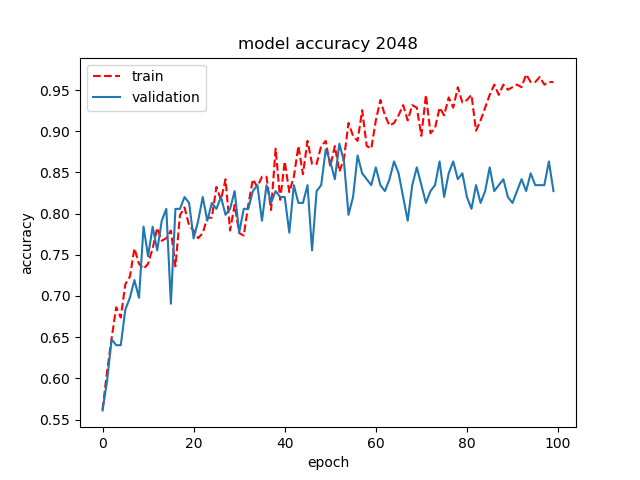}
    \caption{The graph shows the training and validation accuracy obtained between blue and yellow over 100 epochs.}
\label{fig:acc13}
\end{figure}
\begin{figure}
    \centering
    \captionsetup{justification=centering}
    \includegraphics[width=6cm]{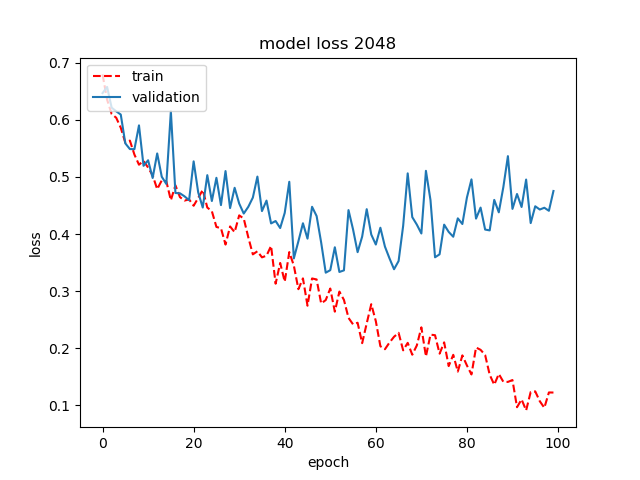}
    \caption{The graph shows the training and validation loss obtained between blue and yellow over 100 epochs.}
\label{fig:loss13}
\end{figure}
\begin{figure}
    \centering
    \captionsetup{justification=centering}
    \includegraphics[width=6cm]{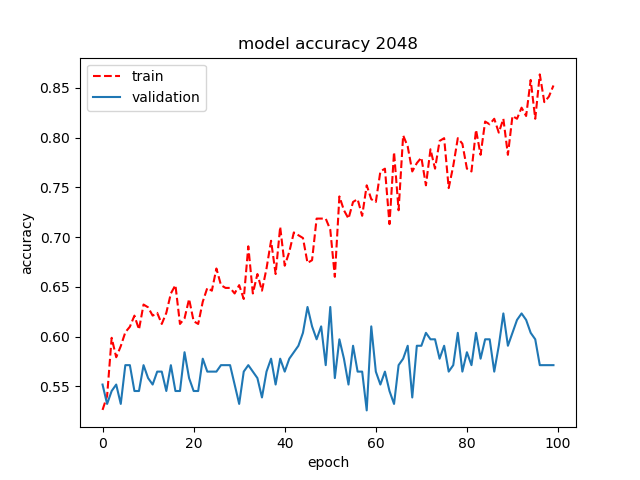}
    \caption{The graph shows the training and validation accuracy obtained between black and yellow over 100 epochs.}
\label{fig:acc23}
\end{figure}
\begin{figure}
    \centering
    \captionsetup{justification=centering}
    \includegraphics[width=6cm]{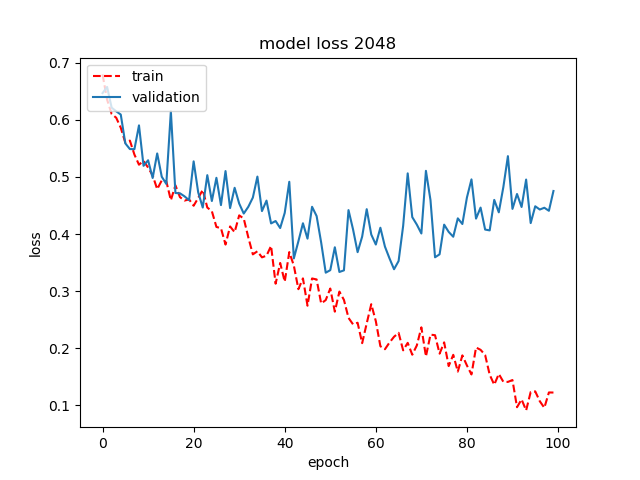}
    \caption{The graph shows the training and validation loss obtained between black and yellow over 100 epochs.}
\label{fig:loss23}
\end{figure}
\subsection{Analysis}
The attention network fit was around 50 epochs for the four-color classifier as in the figure ~\ref{fig:allacc} and figure ~\ref{fig:allloss}. From figure c and figure d it can be analyzed that for two color classifier using red and yellow as the colors (the best case scenario), the fit was around 63 epochs as seen in figure ~\ref{fig:acc03} and ~\ref{fig:loss03}. For other two color classifiers, starting with red and blue, the fit was around  25 epochs as seen in figure ~\ref{fig:acc01} and figure ~\ref{fig:loss01}, for red and black it was around 27 epochs as seen in figure ~\ref{fig:acc02} and figure ~\ref{fig:loss02}, for blue and black it was around 65 epochs as seen in figure ~\ref{fig:acc12} and figure ~\ref{fig:loss12}, for blue and yellow it was around 50 epochs as seen in figure ~\ref{fig:acc13} and figure ~\ref{fig:loss13}.\par
It can be seen in the table attached that using a four color classifier in the neural network the accuracy obtained is 63.75\% whereas while using a two color classifier accuracy jumps to 93.5\% for the same data-set in the same system.
\par
\begin{table}
    \centering
    \captionsetup{justification=centering}
    \caption{Result using two-color classification in attention network between Red and Yellow}
    \begin{tabular}{p{0.2in} p{0.2in} p{0.2in} p{0.3in} p{0.3in} p{0.4in} p{0.4in} p{0.3in} }\toprule
Set&Data Type&Color&Command&Number of data&Correctly Predicted&Falsely predicted&Accur
acy\\ \midrule 
1&Seen&Red&0&50&46&4&92{\%}\\                                                            
2&Seen&Yellow&3&50&45&5&90{\%}\\                                                            
3&Unseen&Red&0&50&48&2&96{\%}\\                                                                                              
4&Unseen&Yellow&3&50&48&2&96{\%}\\ \midrule
&&&&&Average&93.5{\%}\\ \bottomrule
\label{tab:result_lstm}
    \end{tabular}
\end{table}
Now, while comparing the four-color versus two-color classifiers, it can be seen that four color classifiers showed comparatively low accuracy of 63.75\% compared to 93.5\% accuracy of the two color classifier. This is because to classify four signals the base accuracy starts from 25\% whereas for two signals it’s 50\%, there’s 50-50 chance that the signals will be classified correctly in a system classifying two signals.\par
\begin{table}
    \centering
    \captionsetup{justification=centering}
    \caption{Result using four-color classification in attention network}
    \begin{tabular}{p{0.1in} p{0.3in} p{0.3in} p{0.3in} p{0.3in} p{0.3in} p{0.4in} p{0.4in} }\toprule
Set&Data Type&Color&Command&Number of data&Correctly Predicted&Falsely predicted&Accur
acy\\ \midrule 
1&Seen&Red&0&50&39&11&78{\%}\\                                                            
2&Seen&Blue&1&50&38&12&76{\%}\\                                                           
3&Seen&Black&2&50&27&23&54{\%}\\             
4&Seen&Yellow&3&50&36&14&72{\%}\\                                                            
5&Unseen&Red&0&50&32&18&64{\%}\\                                                          
6&Unseen&Blue&1&50&34&16&68{\%}\\                                                         
7&Unseen&Black&2&50&28&22&56{\%}\\                                                            
8&Unseen&Yellow&3&50&29&21&58{\%}\\ \midrule
&&&&&Average&65.75{\%}\\ \bottomrule
\label{tab:result_lstm}
    \end{tabular}
\end{table}

The four-color classifier was then used to feed bluetooth signal to motor-controlled wheelchair prototype with each color associated with the direction the wheelchair would move as seen in the figure~\ref{fig:a2}. To prevent the movement of wheelchair with the slightest deviation of mind, another parameter i.e attention was used, with which the wheelchair moved only with strong EEG signals specifying to a color. 
\par
\section{Conclusion and Future Enhancement}
The Attention based LSTM network classified the two colors with an accuracy of 93.5\% and four colors with an accuracy of 65.75\%.These models are considerable improvements over the random classification of 50\% and 25\% for two class classification problem and four class classification problem respectively. These models can be used to suit many functions like answering of question with yes/no for certain coma patients, directional control of wheelchairs for patients of locomotive disorders, etc., with further improvements in the future. There is enough room for improvement of the models by using techniques like -introducing a better preprocessing method of data for removal of noise, using a more realistic environment and stimuli to reflect real world scenarios, using a more sophisticated EEG sensor with multiple electrodes to get better signals from various regions of the brain used for thought and so on.


\bibliographystyle{IEEEtran}
\bibliography{reference}

\end{document}